\documentclass[aps,prl,amsfonts,floatfix,superscriptaddress,longbibliography,nofootinbib,twocolumn,10pt]{revtex4-2}

\usepackage{graphicx}
\usepackage{amsmath}
\usepackage{physics}
\usepackage{mathtools}
\usepackage{makecell}
\usepackage[hidelinks, colorlinks=true, allcolors=blue]{hyperref}
\usepackage{mathrsfs}

\usepackage[T1]{fontenc}
\renewcommand{\selectlanguage}[1]{}

\begin{document}
\preprint{APS/123-QED}
\title{Hamiltonian Simulation via Stochastic Zassenhaus Expansions}

\author{Joseph Peetz}
\email{peetz@ucla.edu}
\affiliation{Department of Physics and Astronomy, University of California, Los Angeles, California, USA}

\author{Prineha Narang}
\email{prineha@ucla.edu}
\affiliation{Division of Physical Sciences, College of Letters and Science, University of California, Los Angeles, California, USA}
\affiliation{Department of Electrical and Computer Engineering, University of California, Los Angeles, California, USA}

\date{\today}

\begin{abstract}
We introduce stochastic Zassenhaus expansions (SZEs), a class of ancilla-free quantum algorithms for Hamiltonian simulation. These algorithms map nested Zassenhaus formulas onto quantum gates and then employ randomized sampling to minimize circuit depths. Unlike Suzuki-Trotter product formulas, which grow exponentially long with approximation order, the nested commutator structures of SZEs enable high-order formulas for many systems of interest. For a 10-qubit transverse-field Ising model, we construct an 11th-order SZE with 42x fewer CNOTs than the standard 10th-order product formula. Further, we empirically demonstrate regimes where SZEs reduce simulation errors by many orders of magnitude compared to leading algorithms.
\end{abstract}

\maketitle

\textbf{\emph{Introduction:}}  Quantum states $\ket{\psi(t)}$ evolve according to Schrödinger's equation,
\begin{equation}
i \partial_t \ket{\psi(t)}=H\ket{\psi(t)},
\end{equation}
where the Hamiltonian $H$ encodes the energy spectrum of the physical system. For time-independent $H$, the dynamics are described by a matrix exponential,
\begin{equation}
\ket{\psi(t)}=e^{-itH} \ket{\psi(0)}.
\end{equation}
Classically computing this matrix exponential quickly becomes infeasible for large system sizes, and a major motivation for developing large-scale, fault-tolerant quantum computers is to precisely emulate this evolution in polynomial time \cite{wiebe_simulating_2011}. Beyond quantum dynamics \cite{miessen_quantum_2023, ollitrault_molecular_2021}, Hamiltonian simulation is also a critical subroutine for applications in quantum materials science \cite{bauer_quantum_2020, lordi_advances_2021, de_leon_materials_2021}, quantum optimization \cite{farhi_quantum_2014, albash_adiabatic_2018}, and ground state estimation \cite{aspuru-guzik_simulated_2005, poulin_preparing_2009, lin_near-optimal_2020, dong_ground-state_2022, wang_quantum_2023, nam_ground-state_2020}.

Provided a Hamiltonian $H$, quantum algorithms for Hamiltonian simulation map the evolution operator $e^{-itH}$ onto a sequence of hardware-friendly quantum gates. Despite rapid advancements in hardware \cite{kim_evidence_2023, delaney_scalable_2024, google_quantum_ai_and_collaborators_quantum_2024}, the experimental realization of these simulations remains challenging, motivating the co-design of resource-efficient algorithms. Researchers have extensively investigated the Suzuki-Trotter product formulas \cite{lloyd_universal_1996, suzuki_general_1991, childs_theory_2021, childs_nearly_2019}, randomized algorithms like qDRIFT \cite{campbell_random_2019}, and post-Trotter methods such as quantum signal processing \cite{low_optimal_2017} and qubitization \cite{low_hamiltonian_2019-1}. These present a myriad of trade-offs, such that faithfully capturing high-order dynamics typically requires complex algorithmic structures. In particular, high-order product formulas are more precise, but their operator sequences grow exponentially long with the approximation order. This scaling makes high-order simulations prohibitively expensive, often limiting us to fourth- or sixth-order formulas in practice \cite{childs_toward_2018}.

We address this fundamental problem by introducing stochastic Zassenhaus expansions (SZEs), a class of hardware-friendly simulation algorithms which avoid the inherent exponential scaling of product formulas. Rather than repeating the same operators exponentially many times, SZEs leverage nested commutator structures to make high-order simulations feasible for many systems of interest. Accordingly, they enable excellent resource scaling with respect to simulation time, system size, and target precision. 



SZEs combine two mathematical techniques: 1) nested applications of the standard Zassenhaus formula and 2) randomized sampling of higher-order operators, using stochastic approximations similar to qDRIFT \cite{campbell_random_2019}. We introduce these methods separately and then explain how to combine the two, using a simple Hamiltonian for clarity. We then present general, multi-variable error analysis for several important classes of Hamiltonians. Finally, we empirically demonstrate SZEs' advantages over product formulas for the simulation of the transverse-field Ising model.

\textbf{\emph{Nested Zassenhaus Expansions:}}  Consider a Hamiltonian of the form
\begin{equation}
    H = \underbrace{\sum_i a_i P_i}_A + \underbrace{\sum_j b_j P_j}_B,
\end{equation}
where all Pauli strings $P_i$ in $A$ commute, and separately, all $P_j$ in $B$ commute. An important example is the transverse-field Ising model, with $A = - J \sum_i Z_i Z_{i+1}$ and $B = - h \sum_j X_j$. For such systems, time evolution can be modeled by the Zassenhaus formula \cite{magnus_exponential_1954},
\begin{equation} \label{eq: zassenhaus-expansion}
    e^{-i t H} = e^{-i t A} e^{-i t B} e^{\frac{t^2}{2!} [A,B]} e^{i \frac{t^3}{3!} [2B+A,[A,B]]} \cdots.
\end{equation}
Considered the dual of the Baker-Campbell-Hausdorff formula, the Zassenhaus formula is both efficiently computable \cite{casas_efficient_2012} and recursively generalizable for any number of operators \cite{wang_multivariable_2019}.
Truncating to first order and discretizing into $r$ time steps gives the standard Trotterization algorithm for Hamiltonian simulation,
\begin{equation}
     e^{-i t H} = \left(e^{-i t A/r} e^{-i t B/r}\right)^r + \mathcal{O}(t^2/r).
\end{equation}
In order to bound the resulting truncation error by a target precision $\epsilon$, this approach thus requires $r \sim \mathcal{O}(t^2/\epsilon)$ time steps. 

To improve this scaling, we propose mapping $e^{\frac{t^2}{2} [A,B]}$ and higher-order operators in the Zassenhaus formula onto quantum gates. Because the commutator of two Hermitian operators is anti-Hermitian, we can view the second-order exponential as a new time evolution operator with its own Hamiltonian $H_2$. That is, we compute
\begin{equation}
    \frac{1}{2}[A,B] = i \sum_k c_k P_k := -i H_2,
\end{equation}
where $c_k \in \mathbb{R}$, and thus $e^{-i t^2 H_2}$ is unitary. More generally, the nested commutator structures of the Zassenhaus formula guarantee that all of its matrix exponentials indeed remain unitary. Thus, we can continue to compute $H_3 := -\frac{1}{6}[2B+A,[A,B]]$ and higher-order arguments, viewing each as its own Hamiltonian simulation problem. Explicitly, Eq. \eqref{eq: zassenhaus-expansion} becomes
\begin{equation} \label{eq: zassenhaus-H_k_form}
    e^{-i t H} = e^{-i t A} e^{-i t B} \prod_{k=2}^{\infty} e^{-i t^k H_k},
\end{equation}
where $\{H_k\}$ are the $(k-1)$-nested commutators in the Zassenhaus expansion.

From this perspective, we can now choose among many Hamiltonian simulation subroutines to map each exponential onto quantum gates, enabling a rich variety of hybrid algorithms. Here, we consider nested applications of the Zassenhaus formula, recursively generating higher-order sequences of gates. For example, if we suppose that $H' = A' + B'$ with $A'$ and $B'$ composed of internally commuting operators, then
\begin{equation} \label{eq: second-order-zassenhaus-A'-B'}
\begin{aligned}
    e^{-i t^2 H_2} = e^{-i t^2 A'} e^{-i t^2 B'} e^{\frac{t^4}{2} [A', B']} \cdots.
\end{aligned}
\end{equation}
Together with the first-order terms, we generate the following second-order nested Zassenhaus expansion:
\begin{equation}
     e^{-i H t} = \left(e^{-i A t/r} e^{-i B t/r} e^{-i A' t^2/r^2} e^{-i B' t^2/r^2}\right)^r \!\! + \mathcal{O}(t^3/r^2).
\end{equation}
This improves the error scaling compared to Trotterization alone, instead requiring $r \sim \mathcal{O}(t^{3/2} \epsilon^{-1/2})$ time steps for a precision $\epsilon$. This process can be applied repeatedly, so long as we can compute the nested commutators and identify internally commuting subsets. In general, a $k$-th order nested Zassenhaus formula requires $r \sim \mathcal{O}(t^{1+1/k} \epsilon^{-1/k})$ time steps.

\textbf{\emph{Stochastic Approximations:}}  While nested Zassenhaus expansions require fewer time steps $r$, the caveat is an increased number of operators per step. To improve this depth prefactor, we now introduce a randomized sampling scheme for approximating higher-order Zassenhaus operators, with connections to the techniques used by Campbell in Ref. \cite{campbell_random_2019} and Wan et. al. in Ref. \cite{wan_randomized_2022}. This stochastic approach is summarized by the following theorem, where $|H|_1 := \sum_k |c_k|$ denotes the $L_1$ norm of the Pauli decomposition of $H$. The proof is in the Supplementary Materials.

\vspace{4mm}
\noindent \textbf{Theorem}: Consider a Hermitian operator $H$ with Pauli decomposition $H = \sum_k c_k P_k$. For $t \in \mathbb{R}$, the following approximation holds:
\begin{equation} \label{eq: theorem-stochastic-error}
    e^{-i t^m H} = \sum_k p_k e^{-i \theta(t) P'_k} + \mathcal{O}\left(|H|_1^2 t^{2m}\right).
\end{equation}
Here, $p_k := \frac{|c_k|}{|H|_1}$, $P'_k := \textrm{sign}(c_k) P_k$, and $\theta(t) := \sec^{-1}\left(\sqrt{1+(t^m |H|_1)^2}\right)$. \\
\vspace{1mm}

This theorem expands the evolution operator $e^{-i t^m H}$ as a convex combination of unitaries to leading error $\mathcal{O}\left(|H|_1^2 t^{2m}\right)$. In this form, the evolution can then be simulated by randomly sampling the unitaries $\{e^{-i \theta(t) P'_k}\}$ with probabilities $\{p_k\}$ \cite{peetz_simulation_2024, peetz_quantum_2024}. For example, we can approximate the second-order Zassenhaus operator as
\begin{equation}
    e^{-i t^2 H_2} = \sum_k p_k e^{-i \theta(t) P'_k} + \mathcal{O}(t^4).
\end{equation}
This expansion is accurate to first order in the \textit{modified} time parameter $t^2$, thus giving a leading error of $\mathcal{O}(t^4)$. Similarly, the third-order Zassenhaus operator becomes
\begin{equation}
    e^{-i t^3 H_3} = \sum_l p_l e^{-i \phi(t) P'_l} + \mathcal{O}(t^6).
\end{equation}
To implement the full, third-order simulation algorithm, we combine these randomly sampled expansions with the standard first-order Trotter operators:
\begin{equation} \label{eq: sze_13_example}
     e^{-i H t} \approx \!\left(e^{-i A t/r} e^{-i B t/r} \sum_{k,l} p_k p_l e^{-i \theta(t/r) P'_k} e^{-i \phi(t/r) P'_l}\right)^r\!\!\!\!,
\end{equation}
with leading error $\mathcal{O}(t^4/r^3)$. Here, $e^{-i A t/r}$ and $e^{-i B t/r}$ are applied consistently for every time step, whereas $\{e^{-i \theta(t/r) P'_k}\}$ and $\{e^{-i \phi(t/r) P'_l}\}$ are repeatedly sampled according to their probability distributions. In total, this sampling generates only two Pauli rotations per step, a significant reduction compared to implementing the full operator sequences. By treating these higher-order terms probabilistically, we thus reduce the depth prefactor while retaining the favorable scaling in time of $r \sim O(t^{4/3} \epsilon^{-1/3})$.

Notably, for an $m$-th order Zassenhaus operator, the leading error of its stochastic approximation scales as $\mathcal{O}(t^{2m})$. This quadratic difference enables us to implement multiple orders of Zassenhaus operators probabilistically, as opposed to a single order like seen in qDRIFT \cite{campbell_random_2019}. It follows that for a $k$th-order nested Zassenhaus expansion, we can stochastically approximate operators with time orders in the range $[k+1, 2k+1]$. We refer to this class of algorithms as stochastic Zassenhaus expansions (SZEs). Going forward, we use the notation SZE$_{k,p}$ to denote a $k$th-order nested Zassenhaus algorithm with stochastic approximations of orders $[k+1,p]$. For instance, Eq. \eqref{eq: sze_13_example} above represents an SZE$_{1,3}$, whereas the case $k=p$ represents a nested Zassenhaus algorithm with no stochastic components.

\textbf{\emph{Multi-Variable Error Analysis:}} While the standard two-variable Zassenhaus formula in Eq. \eqref{eq: zassenhaus-expansion} captures the main ideas of SZEs, general systems require a multi-variable extension. In Ref. \cite{wang_multivariable_2019}, Wang et. al. define a recursive algorithm to compute the Lie polynomials $W_j(S)$ of the multi-variable Zassenhaus formula,
\begin{equation}
e^{X_1+X_2+\cdots+X_m}=e^{X_1} e^{X_2} \cdots e^{X_m} \prod_{j=2}^{\infty} e^{W_j(S)},
\end{equation}
where $S := \{X_i\}$. For example, the first two Lie polynomials are
\begin{equation}
\begin{aligned}
W_2(S) & =\frac{1}{2} \sum_{1 \leq i<j \leq n}\left[X_j, X_i\right] \quad\; \textrm{and} \\
W_3(S) & =\frac{1}{3} \sum_{1 \leq i<j, k \leq n}\left[\left[X_j, X_i\right], X_k\right] \\
& + \frac{1}{6} \sum_{1 \leq i<j \leq n}\left[\left[X_j, X_i\right], X_i\right]. \\
\end{aligned}
\end{equation}
In general, $W_j(S)$ contains a sum over $(j-1)$-nested commutators of the operators $X_i \in S$.

\renewcommand{\arraystretch}{2.5}
\begin{table}[t!]
    \centering
    \begin{tabular}{c|c|c}
    \textbf{System} & \textbf{PF$_p$} & \textbf{SZE$_{k,p}$} \\ \hline
    \makecell{\textbf{Nearest}\\\textbf{neighbor}} &  $5^{p/2} (nt)^{1+1/p}$ & $g_k (nt)^{1+1/p}$ \\
    \textbf{$j$-local} & $5^{p/2} n^j L |H|_1^{1 / p} t^{1+1 / p}$ & $n^{k(j-1)+1} L |H|_1^{1 / p} t^{1+1 / p}$ \\
    \makecell{\textbf{Electronic}\\\textbf{structure}} & $5^{p/2} n (nt)^{1+1/p}$ &  $n^{k} (nt)^{1+1/p}$ \\
    \makecell{\textbf{Power law}\\\textbf{$(\alpha \mkern-2mu < \mkern-2mu d)$}} & $5^{p/2} n^{3-\frac{\alpha}{d}+\frac{1}{p}\left(2-\frac{\alpha}{d}\right)} t^{1+\frac{1}{p}}$ & $n^{2+k-\frac{\alpha}{d}+\frac{1}{p}\left(2-\frac{\alpha}{d}\right)} t^{1+\frac{1}{p}}$ \\
    \makecell{\textbf{Power law}\\\textbf{$(d \mkern-2mu \leq \mkern-2mu \alpha \mkern-2mu \leq \mkern-2mu 2d)$}} & $5^{p/2} n^{2+\frac{1}{p}} t^{1+\frac{1}{p}}$ & $n^{m+\frac{1}{p}} t^{1+\frac{1}{p}}$ \\
    \makecell{\textbf{Power law}\\\textbf{$(\alpha \mkern-2mu > \mkern-2mu 2d)$}} & $5^{p/2} (n t)^{1+\frac{d}{\alpha-d}+\frac{1}{p}}$ & $(n t)^{1+\frac{d}{\alpha-d}+\frac{1}{p}}$ \\
    \makecell{\textbf{Quasilocal}} &  $5^{p/2} (nt)^{1+1/p}$ & $k (nt)^{1+1/p}$ \\
    \end{tabular}
    \caption{Gate complexities of product formulas (PF) \cite{childs_nearly_2019, childs_theory_2021} vs. stochastic Zassenhaus expansions (SZE), assuming $k \leq p \leq 2k$. We consider nearest-neighbor Hamiltonians on a lattice, $j$-local Hamiltonians with arbitrary connectivity, electronic-structure Hamiltonians, power-law interactions on a $d$-dimensional lattice which decay as $1/x^{\alpha}$, and quasilocal interactions on a lattice which decay exponentially with distance. For nearest-neighbor Hamiltonians, $g_k$ is the number of gates per site, a system-specific quantity, such that the total number of gates per step is $\mathcal{O}(g_k n)$. Further, $L$ is the induced 1-norm \cite{childs_theory_2021}, and we define the parameter $m = \min\left\{k+1, \left\lceil \frac{\alpha/d + 1/p}{\alpha/d + 1/p - 1} \right\rceil \right\}$.}
    \label{table:gate-complexity-abbreviated}
\end{table}

Now, consider the algorithm SZE$_{k,p}$, in which all nested Zassenhaus operators up to order $k$ are implemented directly and all nested operators of orders $[k+1,p]$ are implemented stochastically. The leading error contributions come from 1) the truncation of Zassenhaus order $p+1$ and 2) the stochastic approximation of order $k+1$:
\begin{equation} \label{SZE-error-scaling-general}
    e^{-itH} = \textrm{SZE}_{k,p}(t) + \mathcal{O}\left(\|H'_{p+1}\| t^{p+1} + |H'_{k+1}|_1^2 t^{2(k+1)}\right).
\end{equation}
Here, $H'_{p+1}$ sums over all Zassenhaus polynomials of time order $p+1$, including those in nested expansions, and $\|H'_{p+1}\|$ denotes its spectral norm. From this expression, we see that stochastically approximating up to order $p = 2k+1$ maximizes the leading-order scaling with respect to time. However, the quadratic scaling of the $L_1$ norm $|H'_{k+1}|_1$ can affect the optimal choice of $p$ in practice.

For example, for a nearest-neighbor Hamiltonian expressed as a sum of $\mathcal{O}(n)$ Pauli strings, all nested commutators retain $\mathcal{O}(n)$ operators \cite{childs_nearly_2019}. In Eq. \eqref{SZE-error-scaling-general}, this implies that $||H'_{p+1}|| \sim \mathcal{O}(n)$ but $|H'_{k+1}|_1^2 \sim \mathcal{O}(n^2)$. To preserve the linear scaling in system size, we thus recommend limiting stochastic approximations to order $p \leq 2k$. This ensures that the leading error scales as $\mathcal{O}\left(n t^{p+1}\right)$.

More generally, when $p \leq 2k$, the leading error scales with the spectral norm of nested commutators, allowing us to adapt the extensive error analysis of product formulas \cite{childs_theory_2021}. For a Hamiltonian $H=\sum_{\gamma=1}^{\Gamma} h_\gamma$, Childs et. al. define the commutator scaling factor 
\begin{equation}
    \widetilde{\alpha}_{\text {comm }}=\!\!\!\!\!\!\!\!\!\!\sum_{\gamma_1, \gamma_2, \ldots, \gamma_{p+1}=1}^{\Gamma} \!\!\!\!\!\!\!\!\! \left\|\left[h_{\gamma_{p+1}}, \cdots\left[h_{\gamma_2}, h_{\gamma_1}\right]\right]\right\|.
\end{equation}
While the nested commutators in Zassenhaus formulas differ from those seen in Trotter errors, they are both strictly subsets of the summands above. Thus, for $p \leq 2k$, the leading errors of stochastic Zassenhaus expansions scale as
\begin{equation}
    \left\| \textrm{SZE}_{k,p}(t) - e^{-itH} \right\| = \mathcal{O}\left(\widetilde{\alpha}_{\textrm{comm}} t^{p+1}\right).
\end{equation}

This result determines the asymptotic scaling of the number of time steps $r$, but we also need to carefully consider the number of operators in a single step of an $\textrm{SZE}_{k,p}$ algorithm. In the Supplementary Materials, we derive the full runtime complexity of several important classes of Hamiltonians, presented in abbreviated form in Table \ref{table:gate-complexity-abbreviated}. We find that SZEs offer the greatest advantage for nearest-neighbor Hamiltonians, quasilocal Hamiltonians, and power-law Hamiltonians with decay exponent $\alpha \geq (2-\frac{1}{p})d$ for lattice dimension $d$. For the remaining classes, SZEs are primarily beneficial in specific instances such as $(k,p) = (1,2)$. These findings strengthen the intuition that SZEs are most efficient for geometrically localized Hamiltonians, as they have the most significant commutator cancellations. 

Crucially, SZEs avoid the inherent exponential prefactor $\mathcal{O}(5^{p/2})$ of product formulas, a fundamental limitation which makes high-order formulas prohibitively expensive. SZEs instead leverage nested commutator structures to enable high-order simulations for many systems of interest, although these structures are notably system-specific. For example, for fully saturated nearest-neighbor Hamiltonians, we show in the Supplementary Materials that the SZE prefactor scales as $g_k = \mathcal{O}(3^k)$. Even in this worst-case scenario, however, setting $p = 2k$ ensures that $g_k$ scales favorably over the product formula prefactor of $\mathcal{O}(5^{k})$. Further, we expect realistic systems to have substantially lower costs in practice, which we demonstrate in the following section.

\begin{figure}[t!]
\centering
\includegraphics[width=\columnwidth]{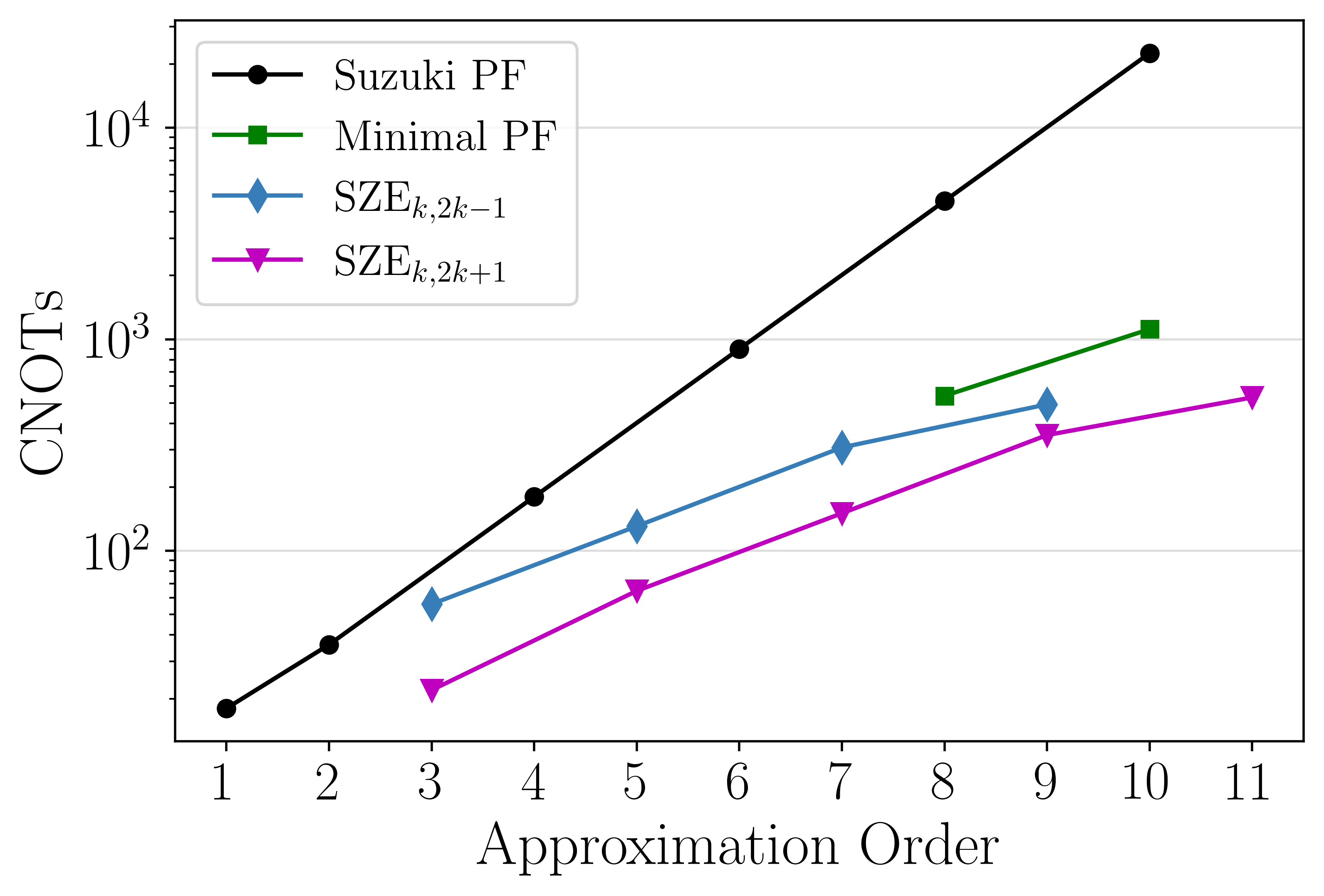} 
\caption{Average number of CNOT gates vs. approximation order $p$ for a tranverse-field Ising model with $n=10$ qubits, before applying compilation algorithms. The stochastic Zassenhaus expansions (SZE) reduce gate counts compared to both the standard Suzuki-Trotter product formulas (PF) and the minimal product formulas \cite{morales_greatly_2022}. For the same $k$-th order nested Zassenhaus formulas, we compare stochastic schemes to order $p=2k-1$ and $p=2k+1$, showing that the extra stochastic terms negligibly increase gate costs.}
\label{figure:TFIM_CNOTs_vs_k}
\end{figure}

\textbf{\emph{Empirical Results:}} We now empirically compare SZEs to product formulas for a benchmark system, the 1D transverse-field Ising model (TFIM):
\begin{equation}
    H = \underbrace{-J \sum_i Z_i Z_{i+1}}_A \underbrace{- h \sum_j X_j}_B.
\end{equation}
Here, $J$ is the exchange interaction parameter, and $h$ quantifies the strength of the transverse magnetic field. This Hamiltonian naturally decomposes into two internally commuting subsets, $A$ and $B$. By computing the nested commutators in the Zassenhaus formula, we can directly apply SZEs to simulate time evolution. For example, we compute the second-order term as
\begin{equation}
    [A,B] = 2i Jh \sum_i (Y_i Z_{i+1} + Z_i Y_{i+1}) = -i H_2.
\end{equation}
This sum partitions into two internally commuting subsets $A'$ and $B'$ corresponding to its even and odd indices, allowing us to simulate $e^{-it^2 H_2}$ via Eq. \eqref{eq: second-order-zassenhaus-A'-B'}. Higher-order Zassenhaus operators can similarly be calculated using Pauli relations, as detailed in the Supplementary Materials. For example, $H_3$ and $H_4$ each partition into four internally commuting subsets, $H_5$ partitions into six, and the nested fourth-order term $H_{2,2} = [A',B']$ partitions into two. All of these retain $\mathcal{O}(n)$ Pauli operators and can thus be efficiently simulated.

\begin{figure}[t!]
\centering
\includegraphics[width=\columnwidth]{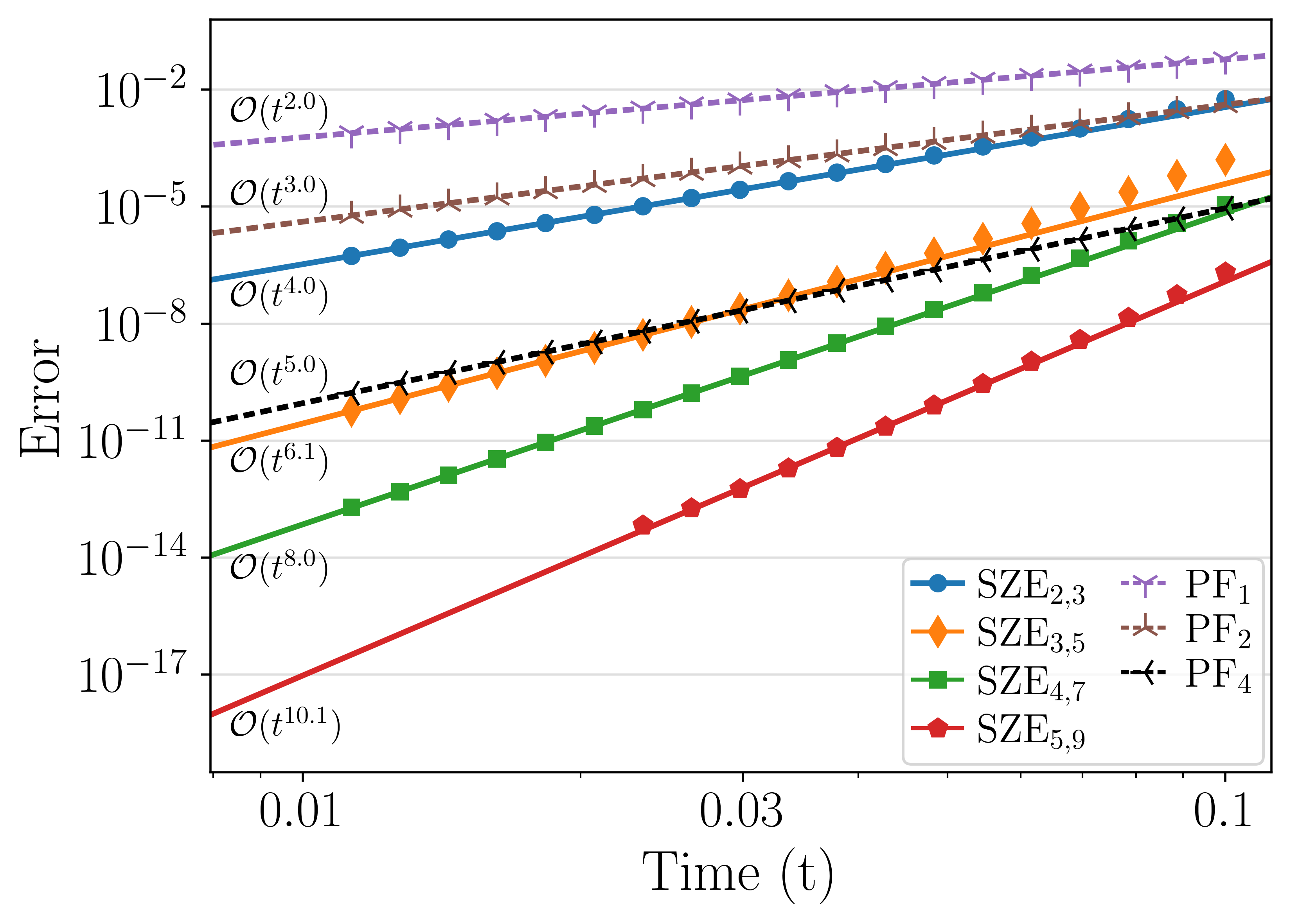} 
\caption{Trace distance error vs. time $t$ for a transverse-field Ising model with $n=10$ qubits and an equal superposition starting state, $\ket{+}^{\otimes n}$. The points are empirically calculated trace distances, and the lines are power-law fits to the smallest $5$ time steps. The labels show each algorithms' asymptotic scaling in time.}
\label{figure:TFIM_trace_dist_vs_time}
\end{figure}

To highlight our approach's performance, we compare the number of CNOT gates required for SZEs and product formulas for several approximation orders $p$ in Figure \ref{figure:TFIM_CNOTs_vs_k}. We see that as $p$ increases, the number of CNOTs grows exponentially for Suzuki-Trotter product formulas but sub-exponentially for SZEs, resulting in an 11th-order SZE with 42x fewer CNOTs than the standard 10th-order product formula. We also compare to recent work on improved product formulas by Morales et. al. \cite{morales_greatly_2022}. They construct 8th- and 10th-order product formulas using a minimum of 15 and 31 2nd-order formulas respectively, in agreement with prior work \cite{yoshida_construction_1990, sofroniou_derivation_2005}. We include these ``Minimal PF" results in Figure \ref{figure:TFIM_CNOTs_vs_k}, showing that our 11th-order SZE requires fewer CNOTs than even the minimal 8th-order product formula.

\begin{figure}[t!]
\centering
\includegraphics[width=\columnwidth]{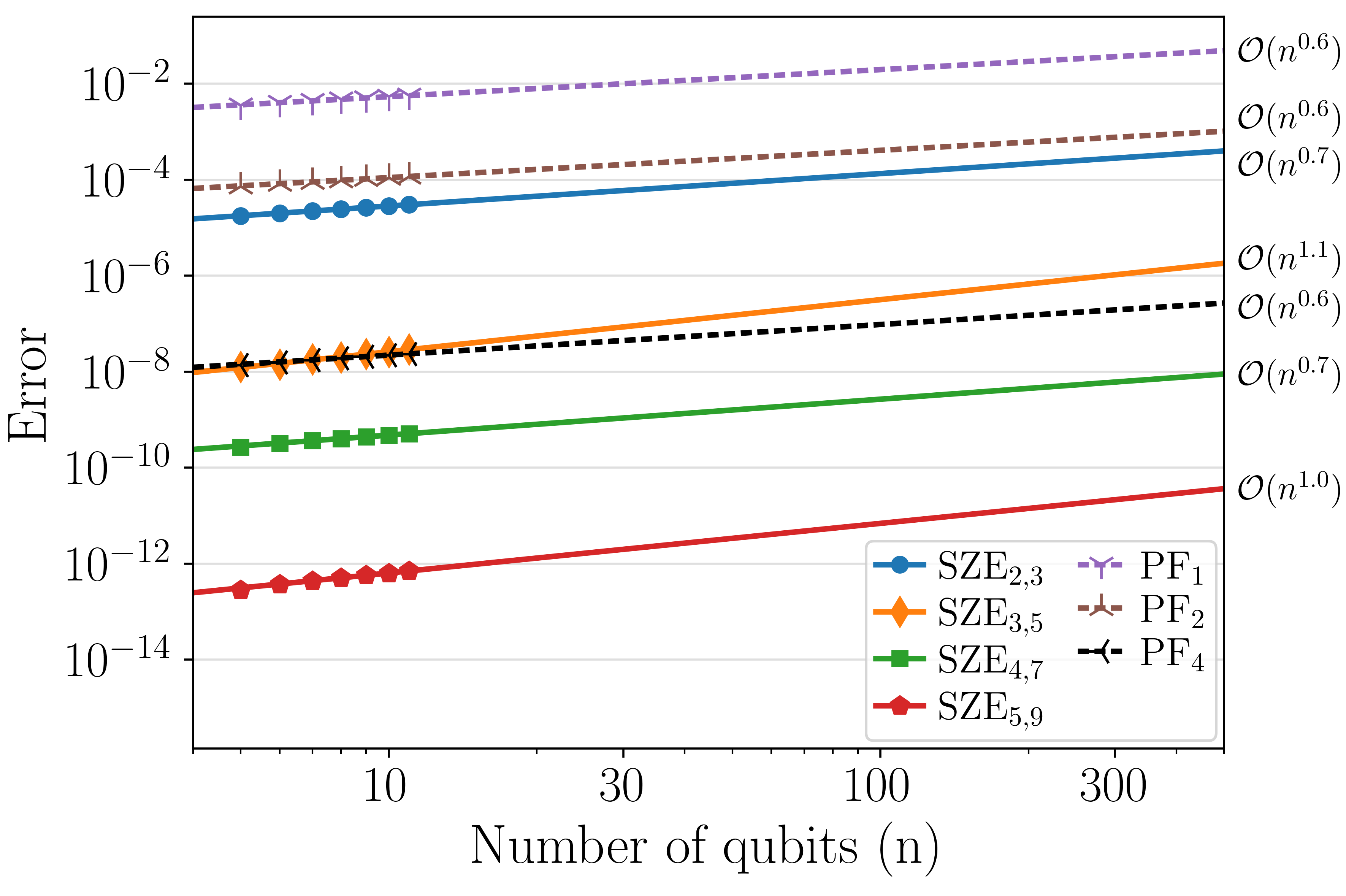} 
\caption{Trace distance error vs. system size $n$ for the transverse-field Ising model. The simulation begins with the equal superposition starting state of $\ket{+}^{\otimes n}$ and evolves for a fixed time step of $t=0.03$. The points are empirically calculated trace distances, and the lines are power-law fits to the largest 5 system sizes. The labels show each algorithms' predicted scaling with respect to system size.}
\label{figure:TFIM_trace_dist_vs_n}
\end{figure}

Next, we empirically compare the trace distance error of SZEs to standard product formulas in Figures \ref{figure:TFIM_trace_dist_vs_time} and \ref{figure:TFIM_trace_dist_vs_n}. In Figure \ref{figure:TFIM_trace_dist_vs_time}, we plot the trace distance error vs. evolution time $t$ for a $10$-qubit TFIM. We see that for the SZE$_{k,p}$ algorithm, the error scales asymptotically as $\mathcal{O}(t^{p+1})$, as predicted. In Figure \ref{figure:TFIM_trace_dist_vs_n}, we plot the trace distance error vs. system size $n$ for a fixed evolution time of $t = 0.03$. We see that the error scales at most linearly in system size, i.e. $\mathcal{O}(n)$. This is consistent with the theoretically predicted scaling of nested commutator structures for nearest-neighbor lattice Hamiltonians \cite{childs_nearly_2019}. Beyond verifying the asymptotic behaviors, in both tests SZEs meaningfully reduce errors by up to several orders of magnitude.

\textbf{\emph{Discussion and Conclusion:}} Stochastic Zassenhaus expansions first map nested Zassenhaus formulas onto sequences of quantum gates up to a truncation order $k$. They then randomly sample higher-order terms up to a stochastic order $p \leq 2k+1$ by expanding them as convex combinations of unitaries. For many systems with geometrically localized interactions, SZEs enable the use of high-order formulas by avoiding the exponentially growing operator sequences of product formulas. SZEs can thus bridge certain trade-offs of leading algorithms, combining the near-term accessibility of product formulas with a resource scaling competitive with optimal algorithms like quantum signal processing \cite{low_optimal_2017}.

While the SZEs investigated here apply a particular recursive scheme, we emphasize that this work enables a rich variety of hybrid algorithms. In particular, in Eq. \eqref{eq: zassenhaus-H_k_form}, one could apply other Hamiltonian simulation subroutines, such as product formulas, to simulate the higher-order time evolution operators $e^{-i t^k H_k}$. Additionally, the qDRIFT-style approximation used in Eq. \eqref{eq: theorem-stochastic-error} can be further enhanced via stochastic combination of unitaries \cite{peetz_quantum_2024}, implementing higher-order remainder terms at the cost of additional repetitions.

Beyond Hamiltonian simulation, Zassenhaus expansions can serve as a broadly useful technique for quantum algorithm design. These techniques can similarly be used to decompose non-unitary matrix exponentials, as seen in applications like imaginary time evolution \cite{wick_properties_1954, lehtovaara_solution_2007, mcardle_variational_2019, motta_determining_2020, nishi_implementation_2021}, ground state preparation \cite{aspuru-guzik_simulated_2005, poulin_preparing_2009, lin_near-optimal_2020, dong_ground-state_2022, wang_quantum_2023, nam_ground-state_2020}, and open quantum system simulation \cite{head-marsden_quantum_2021, schlimgen_quantum_2021, kamakari_digital_2022, ding_simulating_2024}. This approach also naturally enables the fast-forwarding of time evolution \cite{gu_fast-forwarding_2021} for systems whose nested commutators vanish beyond a certain order, allowing for precise simulation in a single time step. Specifically, this speedup occurs for Hamiltonians comprised of operators in a nilpotent Lie algebra, such as the Heisenberg algebra \cite{tanner_nilpotent_1994}. 

\textbf{\emph{Acknowledgements:}}  This work is supported by an NSF CAREER Award under Grant No. NSF-ECCS1944085 and the NSF CNS program under Grant No. 2247007. The authors also gratefully acknowledge Prof. Dong An for the insightful discussions and feedback. This work began while visiting the Institute for Pure and Applied Mathematics, which is supported by the NSF Grant No. DMS-1925919.


%

\clearpage
\onecolumngrid
\section{Supplementary Materials} \label{section: supplementary-materials}

\subsection{Stochastic Approximation}
Here, we explicitly expand matrix exponentials as convex combinations of unitaries, as applied in stochastic Zassenhaus expansions. In this form, each operator can then be approximated via a sampling-based approach, thus lowering the associated gate costs. Specifically, we prove the following theorem stated in the main text.

\vspace{4mm}
\noindent \textbf{Theorem}: Consider a Hermitian operator $H$ with Pauli decomposition $H = \sum_k c_k P_k$. For $t \in \mathbb{R}$, the following approximation holds:
\begin{equation}
    e^{-i t^m H} = \sum_k p_k e^{-i \theta(t) P'_k} + \mathcal{O}\left(|H|_1^2 t^{2m}\right).
\end{equation}
Here, $p_k := \frac{|c_k|}{|H|_1}$, $P'_k := \textrm{sign}(c_k) P_k$, and $\theta(t) := \sec^{-1}\left(\sqrt{1+(t^m |H|_1)^2}\right)$.

\vspace{4mm}
\noindent \textbf{Proof}:  The core idea is to expand the lowest-order terms into a convex combination using a trick similar to that of Ref. \cite[Appendix C]{wan_randomized_2022}:
\begin{equation}
\begin{aligned}
    e^{-i t^m H} & = I - i t^m H - \frac{t^{2m}}{2}H^2 + \mathcal{O}(t^{3m}) \\
    & = I - i t^m \sum_k c_k P_k - \frac{t^{2m}}{2}H^2 + \mathcal{O}(t^{3m}) \\
    & = I - i t^m |H|_1 \sum_k p_k P'_k - \frac{t^{2m}}{2}H^2 + \mathcal{O}(t^{3m})\\
    & = \sum_k p_k \left(I - i t^m |H|_1 P'_k\right) - \frac{t^{2m}}{2}H^2 + \mathcal{O}(t^{3m}),
\end{aligned}
\end{equation}
where $|H|_1 := \sum_k |c_k|$ is the $L_1$ norm of the coefficients and $P'_k \in \pm \{I,X,Y,Z\}^{\otimes n}$ are Pauli strings up to negative signs. Thus, $p_k \!>\! 0 \; \forall \; k$ and $\sum_k p_k = 1$, guaranteeing that $\{p_k\}$ forms a probability distribution. Now, we use the following identity:
\begin{equation}
    I - i x P_k = \sqrt{1 + x^2} e^{-i \theta P_k}, \quad \textrm{with } \theta = \sec^{-1}\left(\sqrt{1+x^2}\right).
\end{equation}
Applying this in the small $t$ limit, we get
\begin{equation}
\begin{aligned}
    e^{-i t^m H} & = \sqrt{1 + t^{2m} |H|_1^2} \sum_k p_k e^{-i \theta(t) P'_k}- \frac{t^{2m}}{2}H^2 + \mathcal{O}(t^{3m}) \\
    & = \sum_k p_k e^{-i \theta(t) P'_k} + \frac{t^{2m}}{2} \left(|H|_1^2 \sum_k p_k e^{-i \theta(t) P'_k} - H^2 \right)+ \mathcal{O}(t^{3m}),
\end{aligned}
\end{equation}
where $\theta(t) := \sec^{-1}\left(\sqrt{1+(t^m |H|_1)^2}\right)$. To understand the leading error, note that the spectral norm of a weighted sum of unitary matrices is bounded above by the $L_1$ norm of their coefficients \cite{choi_norm_2004}. Combined with the submultiplicative property of the spectral norm, it follows that $\left\| H^2 \right\| \leq \left\| H \right\| \left\| H \right\| \leq |H|_1^2$. Thus, the leading error is upper bounded by $|H|_1^2 t^{2m}$, concluding our proof.

\subsubsection{Random-Unitary Sampling Approximation}
A subtle yet important consideration is that sampling the convex decomposition of $e^{-it^m H}$ is itself an approximation which induces error. For an input density matrix $\rho$, the target quantum channel of this evolution is
\begin{equation}
\begin{aligned}
    \mathcal{E}(\rho) & = e^{-i t^m H} \rho e^{i t^m H} \\
    & = \left(\sum_j p_j e^{-i \theta(t) P'_j}\right) \rho \left(\sum_k p_k e^{i \theta(t) P'_k}\right) + \mathcal{O}\left(|H|_1^2 t^{2m}\right) \\
    & = \Bigg(I - i t^m |H|_1 \sum_j p_j P'_j\Bigg) \rho \Bigg(I + i t^m |H|_1 \sum_k p_k P'_k\Bigg) + \mathcal{O}\left(|H|_1^2 t^{2m}\right) \\
    & = \rho - i t^m |H|_1 \sum_j p_j \left( P'_j \rho - \rho P'_j \right) + t^{2m} |H|_1^2 \sum_{j,k} p_j p_k P'_j \rho P'_k + \mathcal{O}\left(|H|_1^2 t^{2m}\right),
\end{aligned}
\end{equation}
where we combined the $\mathcal{O}(t^m)$ terms into an equivalent, unified sum. By randomly implementing a unitary from the convex decomposition of $e^{it^m H}$, we neglect the cross terms in this channel description. Instead, we effectively apply the random-unitary channel
\begin{equation}
\begin{aligned}
    \mathcal{E^{\textrm{RU}}}(\rho) & = \sum_j p_j e^{-i \theta(t) P'_j} \rho e^{i \theta(t) P'_j} \\
    & = \sum_j p_j \Bigg(I - i t^m |H|_1 P'_j\Bigg) \rho \Bigg(I + i t^m |H|_1 p_j P'_j\Bigg) + \mathcal{O}\left(|H|_1^2 t^{2m}\right) \\
    & = \rho - i t^m |H|_1 \sum_j p_j \left( P'_j \rho - \rho P'_j \right) + t^{2m} |H|_1^2 \sum_{j} p_j^2 P'_j \rho P'_j + \mathcal{O}\left(|H|_1^2 t^{2m}\right).
\end{aligned}
\end{equation}
We see that this random-unitary approximation thus neglects the incoherent cross terms. Crucially, however, the leading error remains the same, as seen in the channel difference
\begin{equation}
\begin{aligned}
    \mathcal{E}(\rho) - \mathcal{E^{\textrm{RU}}}(\rho) = \mathcal{O}\left(|H|_1^2 t^{2m}\right).
\end{aligned}
\end{equation}
Further, by mapping the evolution onto a random-unitary channel, no sampling-related variance is introduced, making this an inherently scalable approach \cite{peetz_simulation_2024}. Recent work has shown that we can indeed include these truncated cross terms in our sampling scheme but at the cost of additional query complexity \cite{peetz_quantum_2024}.

\subsection{Worked Example: Transverse-Field Ising Model}
Consider the 1D transverse-field Ising model,
\begin{equation}
    H = \underbrace{-J \sum_i Z_i Z_{i+1}}_A - \underbrace{h \sum_j X_j}_B.
\end{equation}
This Hamiltonian naturally decomposes into two internally commuting subsets $A$ and $B$. Thus, we can directly apply the standard, two-variable Zassenhaus formula. One can derive the following results:
\begin{equation}
\begin{aligned}
    [A,B] & = 2i Jh \sum_i (Y_i Z_{i+1} + Z_i Y_{i+1}) \\
    [2B+A,[A,B]] & = 16Jh^2 \sum_i (Z_i Z_{i+1} - Y_i Y_{i+1}) -4J^2h \sum_{j} (2 X_j + Z_j X_{j+1} Z_{j+2}) + 4Jh^2 (X_1 + X_n).
\end{aligned}
\end{equation}
We now need to identify the operators in the Zassenhaus expansion:
\begin{equation}
\begin{aligned}
    e^{-i H t} & = e^{-i A t} e^{-i B t} e^{\frac{t^2}{2!} [A,B]} e^{i \frac{t^3}{3!} [2B+A,[A,B]]} + O(n t^4) \\
    & = e^{-i A t} e^{-i B t} e^{-i A' t^2} e^{-i B' t^2} e^{-i A'' t^3} e^{-i B'' t^3} e^{-i C'' t^3} e^{-i D'' t^3} + O(n t^4).
\end{aligned}
\end{equation}
From the nested commutators above, we identify:
\begin{equation}
\begin{aligned}
    A' & = -Jh \!\!\!\!\!\!\! \sum_{i \in \{1,3,5,\ldots \}} \!\!\!\!\!\!\! Y_i Z_{i+1} + Z_i Y_{i+1} \\
    B' & = -Jh \!\!\!\!\!\!\! \sum_{i \in \{2,4,6,\ldots \}} \!\!\!\!\!\!\! Y_i Z_{i+1} + Z_i Y_{i+1} \\    
    A'' &= -\frac{8}{3}Jh^2 \sum_i Z_i Z_{i+1} \\
    B'' &= \frac{8}{3}Jh^2 \sum_i Y_i Y_{i+1} \\
    C'' &= \frac{4}{3}J^2h \sum_i X_i + 4Jh^2 (X_1 + X_n) \\
    D'' & = \frac{2}{3}J^2h \sum_i Z_i X_{i+1} Z_{i+2}.
\end{aligned}
\end{equation}
Notice that each of these consists of roughly $n$ internally commuting Pauli strings, so we can directly implement their time evolution operators. 

As is, the leading-order errors thus come from the terms $e^{\frac{t^4}{8} [A', B']}$ and $e^{\frac{-t^4}{24}([[[A, B], A], A]+3[[[A, B], A], B]+3[[[A, B], B], B])}$. Suppose that we then sample the fourth- through sixth-order Zassenhaus terms, including the sixth-order terms created from nested commutators of the operators $\{A'',B'',C'',D''\}$. Then our leading truncation error will scale as $O(n t^7)$, which becomes $O(\frac{n t^7}{r^6})$ after discretizing. Bounding this error to $\epsilon$ thus requires the following number of time steps:
\begin{equation}
    r \sim O\left(\frac{n^{1/6} t^{7/6}}{\epsilon^{1/6}}\right).
\end{equation}
Because each time step involves approximately $8n$ Pauli rotations, in total this algorithm has the following predicted number of gates $d$:
\begin{equation}
\begin{aligned}
    d(\textrm{SZE}_{3,6}) & \sim 8nr \sim 8 n^{7/6} t^{7/6} \epsilon^{-1/6}.
\end{aligned}
\end{equation}
With higher-order expansions, we can continue to improve this complexity at the cost of a growing prefactor. In comparison, a sixth-order product formula achieves the same scaling by repeating the $\mathcal{O}(2n)$ operators of $H$ a total of $2 \cdot 5^2$ times for each step. This results in the following predicted number of gates:
\begin{equation}
\begin{aligned}
    d(\textrm{PF}_6) & \sim 100nr \sim 100 n^{7/6} t^{7/6} \epsilon^{-1/6}.
\end{aligned}
\end{equation}
We see that the stochastic Zassenhaus expansion thus achieves the same asymptotic scaling but with an order of magnitude fewer gates.

\subsection{Error Analysis}
In this section, we derive the runtime complexity for several important classes of Hamiltonians. In doing so, we elucidate the instances where stochastic Zassenhaus expansions are most and least beneficial compared to product formulas. The gate complexity results are summarized in Table \ref{table:gate-complexity-full}, detailing both the number of time steps $r$ and the number of gates per step $\mathcal{G}$. In general, we find that SZEs are most efficient for geometrically localized Hamiltonians, as these have the most significant commutator cancellations.

Much of this analysis relies on the excellent body of existing literature, adapting relevant techniques and results to this algorithm. In particular, we repeatedly make use of Ref. \cite{childs_theory_2021}, which derives the error scaling of $p$-th order product formulas for many classes of interest. Childs et. al. show that for a Hamiltonian $H=\sum_{\gamma=1}^{\Gamma} h_\gamma$, the error of the $p$-th order product formula $\mathscr{S}_p(t)$ scales as
\begin{equation}
    \left\|\mathscr{S}_p(t)-e^{-i t H}\right\| = \mathcal{O}(\widetilde{\alpha}_{\textrm{comm}} t^{p+1}), \quad \textrm{where} \quad \widetilde{\alpha}_{\text {comm }}=\!\!\!\!\!\!\!\!\!\!\sum_{\gamma_1, \gamma_2, \ldots, \gamma_{p+1}=1}^{\Gamma} \!\!\!\!\!\!\!\!\! \left\|\left[h_{\gamma_{p+1}}, \cdots\left[h_{\gamma_2}, h_{\gamma_1}\right]\right]\right\|
\end{equation}
and $\|\cdot\|$ denotes the spectral norm. For a target precision $\epsilon$, it then suffices to discretize into $r = \mathcal{O}(\widetilde{\alpha}_{\textrm{comm}}^{1/p} t^{1+1/p} \epsilon^{-1/p})$ time steps.

While the nested commutators in Zassenhaus formulas generally differ from those seen in Trotter errors, they are both strictly subsets of the summands within $\widetilde{\alpha}_{\text {comm}}$. This implies the inequality $\|H'_{p+1}\| \leq \widetilde{\alpha}_{\text {comm}}$, where $H'_{p+1}$ sums over all Zassenhaus operators of time order $p+1$, including those in nested expansions. This claim is not immediately clear from the recursive formula for $H_k$ \cite[Eq. (2.19)]{casas_efficient_2012}, but Refs. \cite{scholz_note_2006, weyrauch_computing_2009} demonstrate how to expand $H_k$ into a linear combination of left-normal nested commutators. For $p \leq 2k$, the error of the SZE$_{k,p}$ thus scales as
\begin{equation}
    \left\| \textrm{SZE}_{k,p}(t) - e^{-itH} \right\| =  \mathcal{O}\left(\|H'_{p+1}\| t^{p+1}\right) = \mathcal{O}\left(\widetilde{\alpha}_{\textrm{comm}} t^{p+1}\right).
\end{equation}
Accordingly, the asymptotic scaling of the number of time steps $r$ is identical for both product formulas and stochastic Zassenhaus expansions, allowing us to directly apply the results from Ref. \cite{childs_theory_2021}.

The primary challenge then is to determine the number of gates per time step of SZEs for each Hamiltonian class. This involves counting the number of significant operators in general Zassenhaus operators, rather than computing their spectral norm as above. Throughout, we emphasize that the purpose of this analysis is to find asymptotic upper bounds, whereas precise resource estimates will generally require system-specific calculations. Notably, not all summands within $\widetilde{\alpha}_{\text {comm}}$ actually appear, and the group theoretic structure of the Zassenhaus formula further constrains the resulting operators. Nonetheless, this analysis provides important guidance for identifying which types of Hamiltonians are efficiently simulable via SZEs.

\renewcommand{\arraystretch}{2.0}
\begin{table}[t!]
    \centering
    \begin{tabular}{c|c|c|c}
    \textbf{System} & \textbf{Time steps ($r$)} & \textbf{PF$_{p}$ gates ($\mathcal{G}$)} & \textbf{SZE$_{k,p}$ ($\mathcal{G}$)} \\ \hline
    Nearest neighbor & $n^{1/p} t^{1+1/p} \epsilon^{-1/p}$ & $5^{p/2} n$ & $g_k n$ \\ \hline
    $j$-local & $L |H|_1^{1 / p} t^{1+1 / p} \epsilon^{-1/p}$ & $5^{p/2} n^j$ & $n^{k(j-1)+1}$ \\ \hline
    \; Electronic structure 
    \; & $(nt)^{1+1/p} \epsilon^{-1/p}$ & $5^{p/2} n$ &  $n^{k}$ \\ \hline
    \textbf{$1/x^{\alpha} \; (\alpha \mkern-2mu < \mkern-2mu d)$} & \; $n^{1-\frac{\alpha}{d}+\frac{1}{p}\left(2-\frac{\alpha}{d}\right)} t^{1+\frac{1}{p}} \epsilon^{-1/p}$ \; & $5^{p/2} n^2$ & $n^{k+1}$ \\ \hline
    \textbf{$1/x^{\alpha} \; (\alpha \mkern-2mu = \mkern-2mu d)$} & $n^{\frac{1}{p}}(\log n)^{1+\frac{1}{p}} t^{1+\frac{1}{p}} \epsilon^{-1/p}$ & $5^{p/2} n^2$ & $n^{k+1}$\\ \hline
    \textbf{$1/x^{\alpha} \; (d < \alpha \mkern-2mu \leq \mkern-2mu 2d)$} & $n^{\frac{1}{p}} t^{1+\frac{1}{p}} \epsilon^{-1/p}$ & $5^{p/2} n^2$ & $n^m$ \\ \hline
    \textbf{$1/x^{\alpha} \; (\alpha \mkern-2mu > \mkern-2mu 2d)$} & $n^{\frac{1}{p}} t^{1+\frac{1}{p}} \epsilon^{-1/p}$ & \; $5^{p/2} n (n t / \epsilon)^{d /(\alpha-d)}$ \; & \; $n (n t / \epsilon)^{d /(\alpha-d)}$ \; \\ \hline
    Quasilocal & $n^{\frac{1}{p}} t^{1+\frac{1}{p}} \epsilon^{-1/p}$ & \; $5^{p/2} n (\log (n t / \epsilon))^d$ \; & \; $k \log(k) n (\log (n t / \epsilon))^d$ \; \\
    \end{tabular}
    \caption{Gate complexity comparisons for $k \leq p \leq 2k$, where $m = \min\left\{k+1, \left\lceil \frac{\alpha/d + 1/p}{\alpha/d + 1/p - 1} \right\rceil \right\}$ and $L$ is the induced 1-norm $L = \max _l \max _{j_l} \sum_{j_1, \ldots, j_{l-1}, j_{l+1}, \ldots, j_k}\left\|H_{j_1, \ldots, j_k}\right\|$ \cite{childs_theory_2021}. Stochastic Zassenhaus expansions offer the clearest advantage for nearest-neighbor lattice Hamiltonians, rapidly decaying power-law Hamiltonians ($\alpha > 2d$), and quasilocal Hamiltonians, due to the highly localized structures of their nested commutators. They additionally improve the gate complexity for power-law Hamiltonians with $(2-\frac{1}{p})d \leq \alpha \leq 2d$. For the remaining classes, where Zassenhaus sequences grow exponentially with order $k$, SZEs are primarily beneficial in specific instances such as $(k,p) = (1,2)$. Note that in Table \ref{table:gate-complexity-abbreviated}, we omit the polylogarithmic factors included here.}
    \label{table:gate-complexity-full}
\end{table}

\subsubsection{Nearest-Neighbor Hamiltonians}
First, we consider a generalized lattice consisting of $n$ sites with nearest-neighbor interactions, but this analysis naturally extends to systems with constant-range interactions. This class encompasses many systems of interest, including Ising, XY, and Heisenberg models, as well as certain lattice gauge theories like the Toric code. An excellent resource on the performance of product formulas for these lattice systems is Ref. \cite{childs_nearly_2019}. 

As a simple example, in one dimension, we can write a nearest-neighbor Hamiltonian as
\begin{equation}
H=\sum_{j=1}^{n-1} h_{j, j+1},
\end{equation}
where each $h_{j, j+1}$ acts locally on sites $j$ and $j+1$. To gain intuition, consider that a commutator of two such operators is nonzero only when they share a site in common. A given operator will accordingly commute with all but $\mathcal{O}(1)$ others, and thus the Zassenhaus operator $H_2$ retains $\mathcal{O}(n)$ terms. Notably, however, $H_2$ will in general contain next-nearest-neighbor interactions as well:
\begin{equation}
H_2 = \sum_{j=1}^{n-1} h_{j, j+1} + \sum_{j=1}^{n-2} h_{j, j+1, j+2}.
\end{equation}
These 3-local operators arise from the commutators of 2-local terms with a single site in common, such as $[h_{j, j+1}, h_{j+1, j+2}]$. The third-order Zassenhaus operator $H_3$ similarly grows to a 4-local Hamiltonian through a chain of overlapping sites of the form $[h_{j, j+1}, [h_{j+1, j+2}, h_{j+2, j+3}]]$. In general, $H_k$ will contain up to $(k+1)$-local operators:
\begin{equation} \label{eq: nearest-neighbor-Hk}
H_k = \sum_{j=1}^{n-1} h_{j, j+1} + \sum_{j=1}^{n-2} h_{j, j+1, j+2} + \cdots + \sum_{j=1}^{n-k} h_{j, j+1, \ldots, j_k}.
\end{equation}
By similar arguments, for a $d$-dimensional lattice with $n$ sites and nearest-neighbor interactions, $H_k$ will contain $\mathcal{O}(n)$ operators of maximal, $(k+1)$-locality and $\mathcal{O}(kn)$ overall operators of the form $h_{\vec{j}}$. 

A practical consideration is that while Eq. \eqref{eq: nearest-neighbor-Hk} has linearly many terms, implementing a $(k+1)$-local operator often entails decomposing it into the Pauli basis. By definition, the maximally non-local operators $h_{j, j+1, \ldots, j_k}$ contain only non-identity Pauli matrices on each of the sites ${j, j+1, \ldots, j_k}$, thus giving a basis of $3^{k+1}$ Pauli strings. In this worst-case scenario where all such Pauli strings are saturated, we get the bound
\begin{equation}
    g_k = \mathcal{O}(3^k).
\end{equation}
Even in this worst-case scenario, setting $p = 2k$ ensures that SZEs maintain a smaller gate prefactor than product formulas of order $p$, which consist of $\mathcal{O}(5^k)$ Pauli rotations.

We emphasize that the product formulas have a \textit{guaranteed} exponentially scaling prefactor, whereas SZEs have a \textit{worst-case} exponentially scaling prefactor (which is smaller still). For a given system, we recommend empirically computing its number of gates as in Figure \ref{figure:TFIM_CNOTs_vs_k} to ensure accurate resource estimates. For instance, for the TFIM, $H_5$ contains at most 4-local operators, in contrast to the 6-local upper bound. Overall, we expect SZEs to have substantially smaller gate prefactors than product formulas for many systems of interest.

\subsubsection{j-local Hamiltonians}
Next, we consider $j$-local Hamiltonians with arbitrary connectivity, consisting of up to $\mathcal{O}(n^j)$ operators. These can be expressed as
\begin{equation}
H = \sum_{i_1, \ldots, i_j} h_{i_1, \ldots, i_j} = \sum_\gamma h_\gamma,
\end{equation}
where each $h_{i_1, \ldots, i_j}$ acts nontrivially on sites ${i_1, \ldots, i_j}$. Using the abbreviated notation $h_{\gamma}$, Zassenhaus operators $H_k$ consist of commutators of the form
\begin{equation}
H_k \sim \sum_{\gamma_1,\ldots,\gamma_k} \left[h_{\gamma_k}, \ldots,\left[h_{\gamma_2}, h_{\gamma_1}\right]\right].
\end{equation}
Childs et. al. show that these nested commutators are supported on at most $k(j-1)+1$ sites \cite{childs_theory_2021}. The highest-order Zassenhaus operator will thus dominate the gate complexity, requiring $\mathcal{O}(n^{k(j-1)+1})$ gates per time step. Due to this rapid growth with nested order $k$, we expect SZE$_{1,3}$ to perform the best for practical simulations.

\subsubsection{Electronic-Structure Hamiltonians}
In second quantization, electronic-structure Hamiltonians take the form
\begin{equation}
H=\underbrace{\sum_{p, q} h_{p q} a_p^{\dagger} a_q}_{T+U}+\underbrace{\frac{1}{2} \sum_{p, q, r, s} h_{p q r s} a_p^{\dagger} a_q^{\dagger} a_r a_s}_V,
\end{equation}
where $T$ is the kinetic energy, $U$ the nuclear potential energy, $V$ the repulsive electron-electron interaction potential, and $a_p^\dagger$ and $a_p$ are the fermionic raising and lowering operators. In Ref. \cite{babbush_low-depth_2018}, Babbush et. al. reduce this Hamiltonian's number of terms from $\mathcal{O}(N^4)$ to $\mathcal{O}(N^2)$ by changing to a plane wave dual basis, where $N$ is the size of the discrete representation. Using the notation of Ref. \cite{childs_theory_2021}, this takes the abbreviated form
\begin{equation}
H' = \underbrace{\sum_{p, q} t_{p, q} a_p^{\dagger} a_q}_T + \underbrace{\sum_p u_p n_p}_U + \underbrace{\sum_{p, q} v_{p, q} n_p n_q}_V,
\end{equation}
where $n_p = a_p^\dagger a_p$ is the number operator. To compute the scaling of product formula errors, Childs et. al. consider a general second-quantized operator of the form
\begin{equation}
W=\sum_{\vec{p}, \vec{q}, \vec{r}} w_{\vec{p}, \vec{q}, \vec{r}} \underbrace{\cdots\left(a_{p_x}^{\dagger} a_{q_x}\right) \cdots\left(n_{r_y}\right) \cdots}_{\text {at most } l \text { operators}},
\end{equation}
referring to $l$ as the number of layers of $W$. Using commutation rules for second-quantized operators, they show that the commutators $[T,W]$ and $[U,W]$ retain $l$ layers, whereas the commutator $[V,W]$ grows to $l+1$ layers. For electronic-structure Hamiltonians, they show by induction that $(k-1)$-nested commutators of the form $\left[h_{\gamma_{k}}, \cdots\left[h_{\gamma_2}, h_{\gamma_1}\right]\right]$ have $k$ layers. Implementing each Trotter step as in Refs. \cite{ferris_fourier_2014, low_hamiltonian_2019}, they achieve an overall gate complexity of $\tilde{\mathcal{O}}(n^{2+o(1)} t^{1+o(1)})$, including the number of time steps \cite{childs_theory_2021}. Similarly, Zassenhaus exponentials $e^{-i t^k H_k}$ can thus be simulated as a sequence of $\tilde{\mathcal{O}}(n^{k})$ gates. Due to this rapid growth with nested order $k$, we expect SZE$_{1,3}$ to perform the best for practical simulations.

\subsubsection{Power-Law Hamiltonians}
Power-law Hamiltonians consist of 2-local lattice interactions whose strength decays with distance, encompassing important models such as the Coulomb interaction, dipole potentials, and the Van der Waals potential. They can be expressed as
\begin{equation}
    H = \sum_{\vec{j}_1,\vec{j}_{2}} h_{\vec{j}_1,\vec{j}_2}, \quad \textrm{with} \quad \left\|h_{\vec{j}_1,\vec{j}_{2}}\right\| \leq \frac{1}{x_{\vec{j}_1,\vec{j}_{2}}}
\end{equation}
for $\vec{j}_1 \neq \vec{j}_2$. Here, $x_{\vec{i},\vec{j}} := \|\vec{i}-\vec{j}\|_2$ is the Euclidean distance between the sites $\vec{i}$ and $\vec{j}$ on the lattice. Throughout this section, we make use of the following identities from Ref. \cite{childs_theory_2021}:
\begin{subequations}
\begin{align}
    & \left\|e^{-i t H}-e^{-i t \widetilde{H}}\right\| \leq\|H-\widetilde{H}\| t. \label{eq: truncated-H-exp-distance} \\
    & \textrm{For } \alpha = d, \sum_{\vec{j} \in \Lambda \backslash \{\vec{0}\}} \frac{1}{\|\vec{j}\|_2^\alpha}= \mathcal{O}\left(\log n\right). \label{eq: power-law-norm-alpha-d-log} \\
    & \textrm{For } \alpha > d, \sum_{\vec{j} \in \Lambda,\|\vec{j}\|_2 \geq \ell} \frac{1}{\|\vec{j}\|_2^\alpha}=\mathcal{O}\left(\frac{1}{\ell^{\alpha-d}}\right). \label{eq: power-law-norm-above-cutoff}
\end{align}
\end{subequations}

Using Eqs. \eqref{eq: truncated-H-exp-distance} and \eqref{eq: power-law-norm-above-cutoff}, Childs et. al. show that for $\alpha > d$, we can truncate power-law interactions acting beyond a cutoff distance $\ell=\mathcal{O}\left((n t / \epsilon)^{1 /(\alpha-d)}\right)$. This yields a truncated Hamiltonian with $\mathcal{O}(n \ell^d)$ operators and an improved gate complexity specifically when $\mathcal{O}(\ell^d) < \mathcal{O}(n)$. This condition implies $\alpha > 2d$, a class they refer to as ``rapidly decaying" power-law Hamiltonians.

Building on these techniques, we now show how to properly truncate Zassenhaus operators $H_k$ by generalizing the cutoff distance $l$ to a $k$-dimensional cutoff \textit{volume} $V_k$. First, recall that $k$-th order Zassenhaus operators consist of $(k-1)$-nested commutators. For power-law Hamiltonians, each of these nested commutators acts nontrivially on at most $k+1$ sites, growing from an initially $2$-local Hamiltonian $H$ to a $(k+1)$-local Hamiltonian $H_k$:
\begin{equation}
    H_k = \!\!\!\! \sum_{\vec{j}_1,\ldots,\vec{j}_{k+1}} \!\!\!\! h_{\vec{j}_1,\ldots,\vec{j}_{k+1}}, \quad \textrm{with} \quad \left\|h_{\vec{j}_1,\ldots,\vec{j}_{k+1}}\right\| \leq \prod_{i=1}^{k} \frac{1}{(x_{\vec{j}_i,\vec{j}_{i+1}})^\alpha}
\end{equation}
for $\vec{j}_1 \neq \vec{j}_2 \neq \ldots \neq \vec{j}_k$. We now define a truncated Hamiltonian $\widetilde{H_k}$ by removing all interactions beyond a cutoff volume $V_k$, i.e. those satisfying the product constraint $\prod_{i=1}^k x_{\vec{j}_i,\vec{j}_{i+1}} \!>\! V_k$. Practically, we sum over all indices $\vec{j}_1, \ldots, \vec{j}_k \in \Lambda$ and then truncate all final indices in the set
\begin{equation}
    A := \left\{\vec{j}_{k+1} \in \Lambda \;\; \Bigl\vert \;\; 
    x_{\vec{j_k},\vec{j}_{k+1}} > V_k \prod_{i=1}^k \frac{1}{x_{\vec{j}_i,\vec{j}_{i+1}}}\right\}.
\end{equation}
This process defines our truncated Hamiltonian $\widetilde{H_k}$. Using Eqs. \eqref{eq: truncated-H-exp-distance}--\eqref{eq: power-law-norm-above-cutoff}, we then upper bound the resulting truncation error as follows:
\begin{equation} \label{eq: power-law-truncation-error-derivation}
\begin{aligned}
    \left\|e^{-i H_k t^k / r^k }-e^{-i \widetilde{H_k} t^k / r^k }\right\| r & \leq\left\|H_k-\widetilde{H_k}\right\| \frac{t^k}{r^{k-1}} \\
    & \leq \left\| \sum_{\vec{j_1} \in \Lambda} \sum_{\vec{j}_2 \in \Lambda} \frac{1}{(x_{\vec{j}_1,\vec{j}_2})^\alpha} \; \cdots \! \sum_{\vec{j}_k \in \Lambda} \frac{1}{(x_{\vec{j}_{k-1},\vec{j}_k})^\alpha} \sum_{\vec{j}_{k+1} \in A} \frac{1}{(x_{\vec{j}_k,\vec{j}_{k+1}})^\alpha} \right\| \frac{t^{k}}{r^{k-1}} \\
    & = \left\| \sum_{\vec{j_1} \in \Lambda} \sum_{\vec{j}_2 \in \Lambda} \frac{1}{(x_{\vec{j}_1,\vec{j}_2})^\alpha} \; \cdots \! \sum_{\vec{j}_k \in \Lambda} \frac{1}{(x_{\vec{j}_{k-1},\vec{j}_k})^\alpha} \mathcal{O}\left(\frac{(x_{\vec{j}_1,\vec{j}_2} \cdots x_{\vec{j}_{k-1},\vec{j}_k})^{\alpha-d}}{V_k^{\alpha-d}}\right) \right\| \frac{t^{k}}{r^{k-1}} \\
    & = \mathcal{O}\left(\frac{1}{V_k^{\alpha-d}}\right) \left\| \sum_{\vec{j_1} \in \Lambda} \sum_{\vec{j}_2 \in \Lambda} \frac{1}{(x_{\vec{j}_1,\vec{j}_2})^d} \; \cdots \! \sum_{\vec{j}_k \in \Lambda} \frac{1}{(x_{\vec{j}_{k-1},\vec{j}_k})^d} \right\| \frac{t^k}{r^{k-1}} \\
    & = \mathcal{O}\left(\frac{1}{V_k^{\alpha-d}}\right) \left\| \sum_{\vec{i} \in \Lambda} \mathcal{O}((\log n)^{k-1}) \right\| \frac{t^k}{r^{k-1}} \\
    & = \mathcal{O}\left(\frac{t^k n (\log n)^{k-1}}{r^{k-1} V_k^{\alpha-d}}\right).
\end{aligned}
\end{equation}
Thus, we have bounded the error induced by this volume-based truncation. We must now limit this error by our overall target precision of $\epsilon$, which constrains the scaling of the cutoff parameter as follows:
\begin{equation}
    V_k \geq \mathcal{O}\left( \left( \frac{t^k n (\log n)^{k-1}}{r^{k-1} \epsilon} \right)^{1/(\alpha-d)}\right).
\end{equation}
To compute the overall gate complexity of $e^{-i t^k \widetilde{H_k}}$, we now count the number of remaining terms per site $\mathcal{N}_k$ after this truncation. For a $d$-dimensional lattice, let $a$ denote the distance between adjacent sites and let $L = \mathcal{O}(n^{1/d})$ denote the overall length of the lattice, with $L \gg a$. For a given site $\vec{i}$, we count its number of interactions by integrating over $(d-1)$-dimensional surface areas $S_d(x) = \mathcal{O}(x^{d-1})$. For a $(k+1)$-local operator, we repeat this process recursively as follows:
\begin{equation}
\begin{aligned}
    \mathcal{N}_k & \sim \int_a^{L} \cdots  \int_a^{L} \int_a^{V_k / \prod_{i=1}^k x_{\vec{j}_i,\vec{j}_{i+1}}} (x_{\vec{j}_1,\vec{j}_2})^{d-1} \cdots (x_{\vec{j}_{k-1},\vec{j}_k})^{d-1} (x_{\vec{j}_k,\vec{j}_{k+1}})^{d-1} dx_{\vec{j}_1,\vec{j}_2} \cdots dx_{\vec{j}_{k-1},\vec{j}_k} dx_{\vec{j}_k,\vec{j}_{k+1}} \\
    & \sim V_k^d \int_a^{L} \cdots  \int_a^{L} \frac{d x_{\vec{j}_1,\vec{j}_2}}{x_{\vec{j}_1,\vec{j}_2}} \cdots \frac{d x_{\vec{j}_1,\vec{j}_2}}{x_{\vec{j}_{k-1},\vec{j}_k}} \\
    & \sim V_k^d (\log L)^{k-1}
\end{aligned}
\end{equation}
We see that the number of interactions per site accumulates polylogarithmic factors through a similar cancellation process as before. Over all $n$ sites $\vec{i}$, the truncated Zassenhaus operator $\widetilde{H_k}$ thus has $\mathcal{O}(n V_k^d (\log L)^{k-1})$ operators. Thus, the gate complexity per time step of including the $k$-th order Zassenhaus exponential in our simulation scales as
\begin{equation}
\begin{aligned}
    \widetilde{\mathcal{G}_k} & = \mathcal{O}(n V_k^d (\log L)^{k-1}) \\
    & = \mathcal{O}\left( n (\log L)^{k-1} \left( \frac{t^k n (\log n)^{k-1}}{r^{k-1} \epsilon} \right)^{d/(\alpha-d)} \right)
\end{aligned}
\end{equation}
We now substitute $L = \mathcal{O}(n^{1/d})$ and $r = \mathcal{O}(n^{\frac{1}{p}} t^{1+\frac{1}{p}} \epsilon^{-1/p})$ for $\alpha > d$:
\begin{equation}
\begin{aligned}
    \widetilde{\mathcal{G}_k} & = \mathcal{O}\left( n (\log n)^{k-1} \left( \frac{t^k n (\log n)^{k-1}}{n^{\frac{(k-1)}{p}} t^{(k-1)+\frac{(k-1)}{p}} \epsilon^{1-{\frac{(k-1)}{p}}}} \right)^{d/(\alpha-d)} \right) \\
    & = \mathcal{O}\left(n (nt/\epsilon)^{(1-\frac{k-1}{p}) d / (\alpha-d)} (\log n)^{(k-1)(1+d/(\alpha-d))} \right)
\end{aligned}
\end{equation}
For a fixed stochastic order $p$, we see that higher orders $k$ decrease the scaling of $nt/\epsilon$ at the cost of increasing the scaling of the polylog factor $\log n$. 

This analysis is useful only when the truncated operator $\widetilde{H_k}$ yields fewer terms than the starting operator $H_k$. This gives the condition $\widetilde{\mathcal{G}_k} < \mathcal{O}(n^{k+1})$, or equivalently $\alpha/d > \frac{k+1}{k} - \frac{k-1}{kp}$. This is always satisfied for rapidly-decaying power-law Hamiltonians ($\alpha > 2d$) but more nuanced when $d < \alpha \leq 2d$. For example, truncating $H_2 \rightarrow \widetilde{H_2}$ is useful when $\alpha/d > \frac{3}{2} - \frac{1}{2p}$, and truncating $H_3 \rightarrow \widetilde{H_3}$ is useful when $\alpha/d > \frac{4}{3} - \frac{2}{3p}$. For a given $\alpha$ and $k$, we choose the minimum of \{$\widetilde{\mathcal{G}_k}, \mathcal{O}(n^{k+1})\}$ based on these inequalities. Then, the overall gate complexity $\mathcal{G}$ of a single time step of SZE$_{k',p}$ is dictated by the worst-case scaling over all $k \leq k'$:
\begin{equation}
    \mathcal{G} = \max_{k \leq k'} \left\{ \min \left\{\widetilde{\mathcal{G}_k}, \mathcal{O}(n^{k+1})\right\} \right\}.
\end{equation}
For $\alpha/d > 2$, the truncation is always beneficial, and $\mathcal{G}_k$ for the original Hamiltonian ($k=1$) dominates: 
\begin{equation}
    \mathcal{G} = \mathcal{O}\left(n (nt/\epsilon)^{ d / (\alpha-d)} \right)
\end{equation}
For $\frac{3}{2} - \frac{1}{2p} < \alpha/d \leq 2$, we truncate for $k \geq 2$, among which $\widetilde{G_2}$ is greatest:
\begin{equation}
\begin{aligned}
    \mathcal{G} = \max \left\{ \mathcal{O}(n^2), \widetilde{\mathcal{G}_2} \right\} = \begin{cases} 
     \mathcal{O}(n^2), & 2 - \frac{1}{p} \leq \alpha/d \leq 2 \\
     \widetilde{\mathcal{G}_2}, & \frac{3}{2} - \frac{1}{2p} < \alpha/d < 2 - \frac{1}{p}
    \end{cases}
\end{aligned}
\end{equation}
For $\frac{4}{3} - \frac{2}{3p} < \alpha/d \leq \frac{3}{2} - \frac{1}{2p}$, we truncate for $k \geq 3$, among which $\widetilde{G_3}$ is greatest:
\begin{equation}
\begin{aligned}
    \mathcal{G} = \max \left\{ \mathcal{O}(n^2), \mathcal{O}(n^3), \widetilde{\mathcal{G}_3} \right\} = \begin{cases} 
     \mathcal{O}(n^3), & \frac{3}{2} - \frac{1}{p} \leq \alpha/d \leq \frac{3}{2} - \frac{1}{2p} \\
     \widetilde{\mathcal{G}_3}, & \frac{4}{3} - \frac{2}{3p} < \alpha/d < \frac{3}{2} - \frac{1}{p}
    \end{cases}
\end{aligned}
\end{equation}
For $\frac{5}{4} - \frac{3}{4p} < \alpha/d \leq \frac{4}{3} - \frac{2}{3p}$, we truncate for $k \geq 4$, among which $\widetilde{G_4}$ is greatest:
\begin{equation}
\begin{aligned}
    \mathcal{G} = \max \left\{ \mathcal{O}(n^2), \mathcal{O}(n^3), \mathcal{O}(n^4), \widetilde{\mathcal{G}_4} \right\} = \begin{cases} 
    \mathcal{O}(n^4), & \frac{4}{3} - \frac{1}{p} \leq \alpha/d \leq \frac{4}{3} - \frac{2}{3p} \\
    \widetilde{\mathcal{G}_4}, & \frac{5}{4} - \frac{3}{4p} < \alpha/d < \frac{4}{3} - \frac{1}{p}
    \end{cases}
\end{aligned}
\end{equation}
For simplicity, we can coarse grain these splittings and upper bound the complexity as
\begin{equation}
\begin{aligned}
    \mathcal{G} = \begin{cases}
        \mathcal{O}(n^2), \quad & 2 - \frac{1}{p} \leq \alpha/d \leq 2 \\
        \mathcal{O}(n^3), \quad & \frac{3}{2} - \frac{1}{p} \leq \alpha/d < 2 - \frac{1}{p} \\
        \mathcal{O}(n^4), \quad & \frac{4}{3} - \frac{1}{p} \leq \alpha/d < \frac{3}{2} - \frac{1}{p} \\
        \quad \vdots \\
        \mathcal{O}(n^m), \quad & \frac{m}{m-1} - \frac{1}{p} \leq \alpha/d < \frac{m-1}{m-2} - \frac{1}{p} \quad (m > 2)
    \end{cases}
\end{aligned}
\end{equation}
Solving for $m$ and requiring $\mathcal{G} < \mathcal{O}(n^{k+1})$, we get the general complexity
\begin{equation}
    \mathcal{G} = \mathcal{O}(n^m), \quad \textrm{where} \quad m := \min\left\{k+1, \left\lceil \frac{\alpha/d + 1/p}{\alpha/d + 1/p - 1} \right\rceil \right\}.
\end{equation}

It follows that for the highly non-local range of $\alpha \leq d$, we cannot truncate $H_k$ for any order. Thus, their gate complexity per time step will be dominated by the highest-order Zassenhaus operator $H_k$. Because power-law Hamiltonians are $2$-local, $H_k$ will in general require $\mathcal{G} = \mathcal{O}(n^{k+1})$ gates. For this highly non-local case, we thus expect SZE$_{1,3}$ to perform the best for practical simulations. In contrast, the truncation scheme offers a complexity improvement over product formulas when $\mathcal{G} \leq \mathcal{O}(n^2)$, corresponding to the class $\alpha \geq (2-\frac{1}{p})d$.

\subsubsection{Quasilocal Hamiltonians}
We now consider quasilocal systems whose interactions decay exponentially with distance:
\begin{equation}
    H = \sum_{\vec{j}_1,\vec{j}_{2}} h_{\vec{j}_1,\vec{j}_2}, \quad \textrm{with} \quad \left\|h_{\vec{j}_1,\vec{j}_{2}}\right\| \leq e^{-\beta x_{\vec{j}_1,\vec{j}_{2}}}
\end{equation}
for $\beta > 0$. One important example of such a system is the Yukawa potential in nuclear physics. In Ref. \cite{childs_theory_2021}, Childs et. al. show that these can be treated similarly to rapidly decaying power-law Hamiltonians but with a logarithmically small cutoff distance $\ell$. We confirm this scaling by integrating over the truncated interactions as follows, where $\Gamma$ is the upper incomplete Gamma function:
\begin{equation} \label{eq: exp-decay-truncation-error-cases}
\begin{aligned}
    \sum_{\vec{j}_2 \in \Lambda \; | \; x_{\vec{j}_1,\vec{j}_{2}} \geq \ell} \!\!\!\!\!\!\!\!\!\!\! e^{-\beta x_{\vec{j}_1,\vec{j}_{2}}} & \approx \int_{\ell}^{L} e^{-\beta x_{\vec{j}_1,\vec{j}_{2}}} (x_{\vec{j}_1,\vec{j}_2})^{d-1} dx_{\vec{j}_1,\vec{j}_2} \\
    & = - \beta^{-d} \Gamma(d, \beta x) \Big|_{x=\ell}^{x=L}  \\
    & = - \beta^{-d} (d-1)! e^{-\beta x} \sum_{k=0}^{d-1} \frac{(\beta x)^k}{k!} \Big|_{x=\ell}^{x=L} \\
    & = \begin{cases}
        \mathcal{O}(1), & \ell \ll 1 \\
        \mathcal{O}(e^{-\beta \ell} \ell^{d-1}), & \ell > 1.
    \end{cases}
\end{aligned}
\end{equation}
For the case $\ell > 1$, we follow Eq. \eqref{eq: power-law-truncation-error-derivation} to derive the full truncation error of $\mathcal{O}(n t e^{-\beta \ell} \ell^{d-1})$, a modified Lambert W-function. Bounding this error by $\epsilon$ requires a logarithmic cutoff distance of $\ell=\mathcal{O}(\log (n t / \epsilon))$, up to log-of-log corrections. Remarkably, the case $\ell \ll 1$ also shows that truncating \textit{all} interactions ($\ell = 0$) only induces an error of $\mathcal{O}(1)$, a useful result for higher-order Zassenhaus terms.

As with power-law systems, the Zassenhaus operator $H_k$ will act on up to $k+1$ sites,
\begin{equation}
    H_k = \!\!\!\! \sum_{\vec{j}_1,\ldots,\vec{j}_{k+1}} \!\!\!\! h_{\vec{j}_1,\ldots,\vec{j}_{k+1}}, \quad \textrm{with} \quad \left\|h_{\vec{j}_1,\ldots,\vec{j}_{k+1}}\right\| \leq \prod_{i=1}^{k} e^{-\beta x_{\vec{j}_i,\vec{j}_{i+1}}}.
\end{equation}
Rather than using cutoff ``volumes" as before, this extremely fast decay lets us truncate all interactions between more than two sites. That is, we define the truncated Hamiltonian $\widetilde{H_k}$ as only the 2-local interactions of $H_k$ within the cutoff distance $\ell_k$. Using Eq. \eqref{eq: exp-decay-truncation-error-cases}, we bound the resulting error as
\begin{equation}
    \left\|H_k-\widetilde{H_k}\right\| \leq  \left\|\sum_{\vec{j}_1 \in \Lambda} \underbrace{\sum_{\substack{\vec{j}_2 \in \Lambda \; | \\ x_{\vec{j}_1,\vec{j}_{2}} \geq \ell_k}} \!\! e^{-\beta x_{\vec{j}_1,\vec{j}_{2}}}}_{\mathcal{O}(e^{-\beta \ell_k} \ell_k^{d-1})} \underbrace{\sum_{\vec{j}_3 \in \Lambda} e^{-\beta x_{\vec{j}_2,\vec{j}_{3}}}}_{\mathcal{O}(1)} \cdots \underbrace{\sum_{\vec{j}_{k+1} \in \Lambda} e^{-\beta x_{\vec{j}_k,\vec{j}_{k+1}}}}_{\mathcal{O}(1)} \right\| = \mathcal{O}\left(n e^{-\beta \ell_k} \ell_k^{d-1} \right).
\end{equation}
Following Eq. \eqref{eq: power-law-truncation-error-derivation}, the full truncation error is then $\mathcal{O}\left(\frac{n t^k}{r^{k-1}} e^{-\beta \ell_k} \ell_k^{d-1}\right)$. Bounding this by $\epsilon$ and substituting $r = \mathcal{O}(n^{\frac{1}{p}} t^{1+\frac{1}{p}} \epsilon^{-1/p})$ gives
\begin{equation}
\begin{aligned}
    \ell_k & \sim  \log (\frac{n t^k}{\epsilon r^{k-1}}) \sim \log (\frac{n t^k}{n^{\frac{(k-1)}{p}} t^{(k-1)+\frac{(k-1)}{p}} \epsilon^{1-{\frac{(k-1)}{p}}}}) = \left(1-{\frac{(k-1)}{p}}\right)\log (n t / \epsilon)
\end{aligned}
\end{equation}
up to log-of-log factors. We see that the truncation distance $\ell_k$ decreases with higher orders, but for simplicity we upper bound all by $\ell_k \leq \mathcal{O}(\log (n t / \epsilon))$. Implementing the operator $e^{-i H_k t^k / r^k}$ thus requires $\mathcal{O}(n (\log (n t / \epsilon))^d)$ gates.

For a given expansion SZE$_{k,p}$, we need to consider not only $H_k$ but all nested Zassenhaus operators with time order $k$ or less. This amounts to counting the number of divisors $d(m)$ for all $m \leq k$, a quantity which Dirichlet bounded as $\mathcal{O}(k \log(k))$ \cite[Theorem 3.3]{apostol_introduction_1998}. Thus, for quasilocal Hamiltonians on a lattice, the total gate cost per time step of the SZE$_{k,p}$ algorithm scales as
\begin{equation}
     \mathcal{G} = \mathcal{O} \left(k \log(k) n (\log (n t / \epsilon))^d \right).
\end{equation}

\end{document}